\documentclass[prl,superscriptaddress,twocolumn,letterpaper]{revtex4}
\usepackage{bm,graphicx,graphics,amsmath,amssymb,bm,epsfig,color}
\usepackage{euscript,tabularx}
\usepackage{textcomp}
\usepackage{gensymb}
\usepackage{longtable}

\begin{document}
\bibliographystyle{apsrev}

 \title{Bistability of vortex core dynamics in a single
   perpendicularly magnetized nano-disk}

\author{G. de Loubens}
\thanks{Corresponding author: gregoire.deloubens@cea.fr}
\affiliation{Service de Physique de l'\'Etat Condens\'e (CNRS URA 2464), CEA Saclay, 91191 Gif-sur-Yvette, France}

\author{A. Riegler}
\affiliation{Physikalisches Institut (EP3), Universit\"at W\"urzburg, 97074
  W\"urzburg, Germany}

\author{B. Pigeau}
\affiliation{Service de Physique de l'\'Etat Condens\'e (CNRS URA 2464), CEA Saclay, 91191 Gif-sur-Yvette, France}

\author{F. Lochner}
\affiliation{Physikalisches Institut (EP3), Universit\"at W\"urzburg, 97074
  W\"urzburg, Germany}

\author{F. Boust}
\affiliation{ONERA, Chemin de la Huni\`ere, 91761 Palaiseau, France}

\author{K.~Y. Guslienko}
\affiliation{Department of Materials Physics, The University of the
  Basque Country, 20080 San Sebastian, Spain}

\author{H. Hurdequint}
\affiliation{Laboratoire de Physique des Solides, Universit\'e Paris-Sud, 91405 Orsay, France}

\author{L.~W. Molenkamp}
\affiliation{Physikalisches Institut (EP3), Universit\"at W\"urzburg, 97074
  W\"urzburg, Germany}

\author{G. Schmidt}
\thanks{Present address: Institut f\"ur Physik,
  Martin-Luther-Universit\"at, Halle Wittenberg, 06099 Halle, Germany}
\affiliation{Physikalisches Institut (EP3), Universit\"at W\"urzburg, 97074
  W\"urzburg, Germany}

\author{A.~N. Slavin}
\affiliation{Department of Physics, Oakland University, Michigan 48309, USA}

\author{V.~S. Tiberkevich}
\affiliation{Department of Physics, Oakland University, Michigan 48309, USA}

\author{N. Vukadinovic}
\affiliation{Dassault Aviation, 78 quai Marcel Dassault, 92552 Saint-Cloud, France}

\author{O. Klein}
\affiliation{Service de Physique de l'\'Etat Condens\'e (CNRS URA 2464), CEA Saclay, 91191 Gif-sur-Yvette, France}

\date{\today}

\begin{abstract}
  Microwave spectroscopy of individual vortex-state magnetic
  nano-disks in a perpendicular bias magnetic field, $H$, is performed
  using a magnetic resonance force microscope (MRFM). It reveals the
  splitting induced by $H$ on the gyrotropic frequency of the vortex
  core rotation related to the existence of the two stable polarities
  of the core. This splitting enables spectroscopic detection of the
  core polarity. The bistability extends up to a large negative
  (antiparallel to the core) value of the bias magnetic field $H_r$,
  at which the core polarity is reversed. The difference between the
  frequencies of the two stable rotational modes corresponding to each
  core polarity is proportional to $H$ and to the ratio of the disk
  thickness to its radius. Simple analytic theory in combination with
  micromagnetic simulations give quantitative description of the
  observed bistable dynamics.
\end{abstract}

\maketitle

\begin{figure}
  \includegraphics[width=\columnwidth]{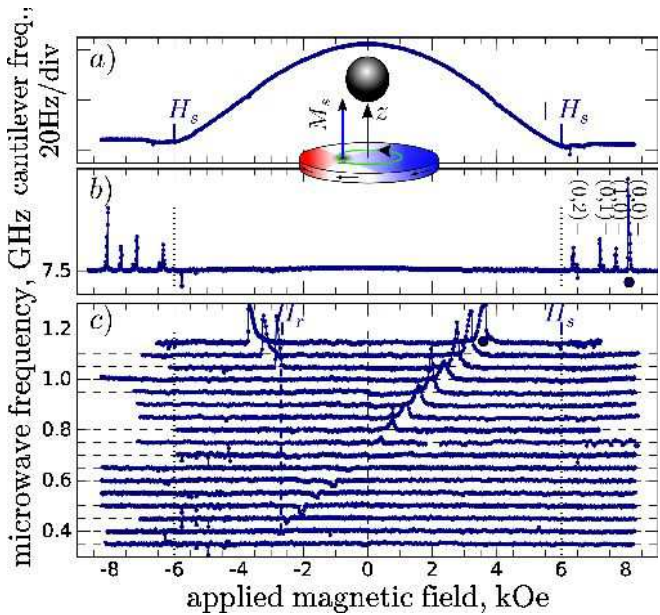}
  \caption{(Color online) (a) The cantilever frequency, proportional
    to $M_z$, the vertical component of the static magnetization of
    the $R_1 = 130$~nm NiMnSb nano-disk, as a function of the
    perpendicular bias magnetic field $H$; (inset) Detection scheme:
    the vortex gyrotropic mode is sensed through the dipolar force
    induced on the spherical Fe probe; (b) Nano-disk excitation
    spectrum in the saturated regime ($|H|> H_s$) recorded at 7.5 GHz;
    (c) Excitation spectra in the vortex (unsaturated) state ($|H|<
    H_s$).}
  \label{spec_130}
\end{figure}

Magnetic vortices are singular topological states found in the
equilibrium magnetic configuration of sub-micron size ferromagnetic
dots \cite{cowburn99,shinjo00}. In a certain range of dot aspect
ratios (ratio $\beta=t/R$ of the dot thickness $t$ to its radius $R$)
the equilibrium ground state of the static magnetization consists of
the curling in-plane magnetization and a nanometer size core of the
out-of-plane magnetization at the dot center. The magnetization of the
vortex core can point either up or down, both polarities $p=\pm 1$
being degenerate at zero field. This bi-stable property of magnetic
vortices, as well as the switching from one polarity to the other,
have been intensively studied in the past few years because of their
possible applications in magnetic storage devices
\cite{waeyenberge06,curcic08,yamada07,guslienko08b}. It has already
been established : (i) that the lowest excitation mode of the vortex
state is the gyrotropic mode corresponding to a rotation of the vortex
core about the dot center, (ii) that the frequency of this mode is
linearly proportional the dot aspect ratio $\beta$ \cite{guslienko02},
and (iii) that the sense of gyration of the vortex core is determined
by a right-hand rule to the core polarity \cite{curcic08}.

In this Letter, we report that by using the exquisitely sensitive
method of magnetic resonance force microscopy (MRFM) \cite{klein08},
we were able to observe bistability of the vortex core dynamics in a
\emph{single} magnetic disk subjected to a perpendicular bias magnetic
field, that was varied in a wide range from positive (parallel to the
vortex core) to negative (antiparallel to the vortex core) values. We
demonstrate that in a certain range of the bias field magnitudes there
are two stable gyrotropic modes of the vortex core rotation having
different frequencies and opposite circular polarizations, and
corresponding to opposite orientations of the vortex core relative to
the direction of the bias magnetic field. The difference in
frequencies of these two stable gyrotropic modes is proportional to
the magnitude of the applied bias field, $H$, and, also, to the dot
aspect ratio $\beta$. We believe that this effect might be important
for the development of novel magnetic memory elements. It allows one
to determine the polarity of the vortex core by measuring the
frequency of the resonance absorption in a nano-disk subjected to a
perpendicular field, which is substantially easier to implement than
the detection of the sense of signal circular polarization, as it was
suggested in \cite{curcic08}.

Our experiments were performed at room temperature on individual
nano-disks of the thickness $t=43.8$ nm and two different radii ($R_1=
130$ and $R_2= 520$~nm). The nano-disks were patterned from a film of
NiMnSb, a soft conductive magnetic material having very low magnetic
losses (typical Gilbert damping constant is $ \alpha=0.002-0.003$),
epitaxially grown on an InP(001) substrate \cite{bach03}. A reference
sample of a continuous film was cut out of the same film for
characterization purposes. A 50~nm thick Si$_3$N$_4$ cap layer was
deposited on top of the disks for protection, and a 300~nm thick Au
broadband strip-line microwave antenna was evaporated on top of the
disks. This antenna generates a linearly polarized microwave magnetic
field oriented in-plane perpendicularly to the stripe direction. The
disk samples were placed in the uniform external bias magnetic field
oriented perpendicular to the disk plane and having magnitude that was
continuously varied from $-10$ to $+10$~kOe.

The detection scheme of MRFM is inspired by magnetic force microscopy
(MFM) \cite{klein08}. It consists of an ultra-soft cantilever with a
800~nm diameter sphere of amorphous Fe (with 3\% Si) glued to its
apex. The magnetization curve of this probe is typical of a soft Fe
sphere (coercitivity $<10$~G). Its magnetic moment always follows the
direction of the applied field $H$ and senses a dipolar force
proportional to the perpendicular component $M_z$ of the magnetization
of the nano-disk. MRFM spectroscopy is achieved by placing the center
of the sphere above the center of the nano-disk. A ferromagnetic
resonance (FMR) spectrum is obtained by recording the vibration
amplitude of the cantilever as a function of $H$ at constant microwave
excitation frequency, that is switched ON and OFF at the cantilever
resonance frequency.  The MRFM signal originates from the diminution
of $M_z$ of the nano-disk produced by the absorbtion of the microwave
field \cite{klein08}. During the scan, the cantilever resonance
frequency is also recorded (see Fig.\ref{spec_130}a). This enables
static magnetometry of the sample. The saturation field of the
$R_1=130$~nm nano-disk is $H_s = \pm 6$ kOe, which marks the region of
the maximum radial susceptibility.

Fig.\ref{spec_130}b shows the FMR spectrum of the smaller nano-disk at
the frequency of 7.5~GHz. Above the saturation field $H_s$, the
excitation spectrum of the disk consists of a series of peaks
corresponding to the confined dipole-exchange spin-wave modes of the
disk. Their quantized resonance frequencies (resonance fields) are
given by Eq.(1) from \cite{kakazei04}.  These modes are labeled with a
pair $(l,m)$ of azimutal and radial mode indices \cite{klein08}.  The
lowest $(0,0)$ (and the most spatially uniform) spin wave mode of the
disk is situated at 8.1~kOe and is marked by a blue circle symbol. The
frequency of this lowest quasi-uniform Kittel mode $\omega_K$ can be
approximately described as:
\begin{equation}
  \omega_K\left(H\right) = \gamma \left\{ H - H_K\right\},
\label{uniform}
\end{equation}
where $H_K = 4 \pi M_s \left (N_{zz} - N_{rr}\right
)-H_\text{A}$. Here $M_s$ is the saturation magnetization, $N_{ii}$
are the diagonal elements of the effective demagnetization tensor of
the uniformly magnetized disk, $H_A$ is a uniaxial perpendicular
anisotropy field (easy-plane for $H_A < 0$), and $\gamma$ is the
gyromagnetic ratio. All the parameters used in Eq.~(\ref{uniform})
were measured independently using the reference film sample by means
of standard magnetometry and cavity-FMR techniques (results are
presented in table \ref{ref}). The calculated position of the Kittel
mode (as well as the positions of the higher modes calculated using
Eq.~(1) from \cite{kakazei04}) are shown near the corresponding index
pairs in Fig.\ref{spec_130}b.  This calculation did not use any
adjustable parameters and took into account the stray field of the
cantilever probe (around 500 Oe) \cite{klein08}. The analytically
calculated values of $H_K$ for the smaller and larger disks are $H_K
\approx 5.7$~kOe and $H_K \approx 8.0$~kOe, respectively.

\begin{table}
  \caption{\label{ref} Physical parameters of the reference NiMnSb layer.}
\begin{ruledtabular}
\begin{tabular}{c c c c}
$4 \pi $Ms (G) & $\gamma$ (rad.s$^{-1}$.G$^{-1}$) & $H_A$ (G) & $\alpha$ \\
$6.9 \times 10^{3}$ & $1.8 \times 10^7$
& $-1.85 \times 10^{3}$ & $2.3 \times 10^{-3}$\\
\end{tabular}
\end{ruledtabular}
\end{table}

When the external magnetic field $H$ is reduced below the saturation
field $H_s$, a single magnetic vortex is formed inside the nano-disk
(which can be visualized by standard MFM). Fig.\ref{spec_130}c shows
the dynamical behavior of the vortex-state disk (unsaturated
regime). The absorption peak observed in the spectra corresponds to
the excitation of the gyrotropic mode in the disk (see inset of
Fig.\ref{spec_130}a) \cite{novosad05}.  It produces a mechanical
signal mainly because the contribution of the vortex core to $M_z$
diminishes when it is excited \cite{guslienko08b}. Such signal is
sensitive to the polarity of the vortex core: it is positive when the
sphere magnetization and the vortex core are parallel and negative in
the opposite case. Moreover, due to the field dependence of the
magnetic moment of the sphere, the amplitude of the MRFM signal
decreases as $|H|$ is reduced to zero.

The baseline of each spectrum in Fig.\ref{spec_130}c is set at the
excitation frequency.  The spectrum for 1.15~GHz has an intense peak
at around $+3.5$~kOe (blue circle) corresponding to the excitation of
the gyrotropic mode of the vortex core, as well as its mirror image at
$-3.5$~kOe.  The gyrotropic frequency decreases with the bias field
$H$, and in the zero field ($H=0$) has the value of 0.7~GHz.  In the
region of negative bias fields $H < 0$ the mode continues to exist,
but the corresponding resonance peak has a negative amplitude, which
indicates that now the direction of the bias field and the amorphous
sphere magnetization is opposite to that of the vortex core.  The
gyrotropic frequency continues to decrease with the same slope until
the bias field value of $H_{r}= -2.8$~kOe, where the resonance peak
abruptly changes its sign, frequency, and slope, which indicates
reversal of the vortex core polarity.  Below the switching field $H <
H_r$ the gyrotropic frequency of the inverted vortex core increases
like it did in positive bias fields before the core reversal. It is
important to note, that upon \emph{increasing} the bias field, the
gyrotropic mode persists at reduced frequency until the symmetric
reversal field of $H_{r} =+2.8$~kOe is reached, and the new reversal
of the vortex core polarity takes place.

\begin{figure}
\includegraphics[width=\columnwidth]{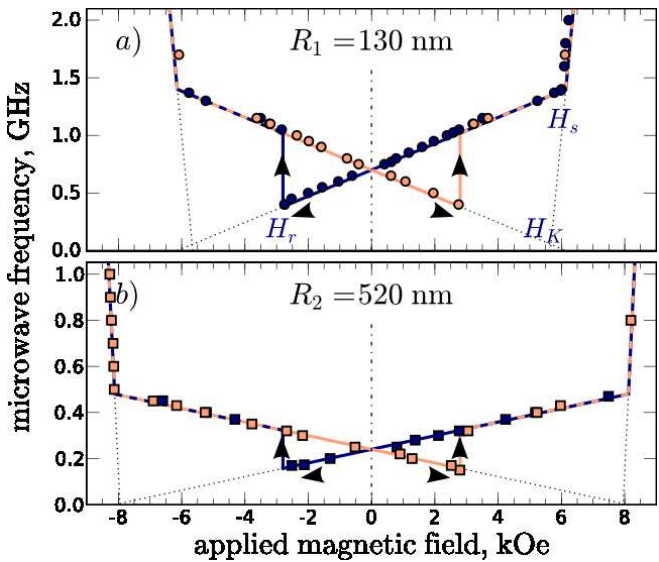}
\caption{(Color online) Frequency of the lowest excitation mode in a
  magnetic disk as a function of the perpendicular bias magnetic field
  for two sizes: (a) $R_1 = 130$~nm, (b) $R_2 = 520$~nm. Dark blue
  (light red) dots are experimental points corresponding to the field
  variation starting from large positive (negative) values.  Solid
  lines are theoretical predictions from Eq.~(\ref{uniform}) for $|H|
  > H_{s}$ and from Eq.~(\ref{slavin}) for $|H|<H_s.$}
\label{vortex}
\end{figure}

The complete phase diagrams demonstrating the dynamic hysteresis loops
of the gyrotropic mode frequency in a perpendicular bias magnetic
field for two disk sizes are presented in Fig.\ref{vortex}. The dark
blue dots correspond to the case when the bias field is reduced
starting from large positive values, while the light red dots
correspond to the opposite case, when the field is varied starting
from large negative values.

Several important features emerge in these phase diagrams: (i) For
$|H|> H_{s}$ the excitation frequency of the lowest mode in the disk
is well described by the Kittel's Eq.~(\ref{uniform}), and at $|H|
=H_{s}$ the field slope of the mode frequency changes abruptly, while
the expected discontinuity of the mode frequency is not seen in the
experiment; (ii) In the interval $| H | < | H_r |$ each value of the
bias field (except $H=0$) corresponds to two different gyrotropic mode
frequencies corresponding to two opposite orientations of the vortex
core relative to the bias field, and the difference between these
frequencies increases with the bias field $H$ and with the disk aspect
ratio $\beta$; (iii) The magnitude of the bias field $H=H_{r}$, at
which the reversal of the vortex core polarity occurs, seems to be
approximately the same for both studied disks.

The frequency of the gyrotropic mode in the external perpendicular
bias magnetic field $H$ can be calculated as a ratio
$\omega_{G}(H)=\kappa(H)/G(H)$ of the field-dependent vortex stiffness
$\kappa(H)$ to the field-dependent magnitude $G(H)$ of the
$z$-component of the vortex gyrovector using the method of Thiele's
equation that was used in \cite{guslienko02} to calculate this
frequency in the case when $H=0$. According to the general definition
given in \cite{guslienko02} (see Eq.~(2) in \cite{guslienko02}) when
the perpendicular bias field is introduced we can write the vortex
gyrovector as $G(H)=G(0)(1-p\cos\theta)$, where $p=\pm1$ is the vortex
core polarity, $G\left(0\right)$ is the gyrovector in zero bias field
defined in \cite{guslienko02}, and $\theta$ is the polar angle of the
static magnetization at the disk lateral boundary. For a sufficiently
large disk radius (much larger than the radius of the vortex core) is
is possible to estimate $\theta$ from the standard electrodynamic
boundary conditions at the vertical boundaries of the disk as
$\cos\theta=H/H_{s}$.

The field-dependent vortex stiffness $\kappa(H)$ is found from the
assumption that the main contribution to the vortex energy comes from
the dipolar interaction of the volume magnetostatic charges created by
the in-plane magnetization component of the shifted vortex outside the
core \cite{guslienko02}. In this region the in-plane magnetization
components depend of the direction of the external bias field as
$\sin\theta$, and the stiffness $\kappa(H)$, that is proportional to
the square of the radial derivative of the in-plane magnetization, is
expressed as $\kappa\left(H\right) =
\kappa\left(0\right)\sin^{2}\theta$, where $\kappa(0)$ is the vortex
stiffness in the zero bias field defined in \cite{guslienko02}.  Thus,
we can write the explicit expression for the gyrotropic mode frequency
in the perpendicular bias field as
\begin{equation}
  \omega_G(H) = \omega_G(0) \left \{ 1 + p \frac{H}{H_s} \right \},
\label{slavin}
\end{equation}
where $\omega_G(0)$ is the gyrotropic mode frequency in the zero bias
field defined by Eq.~(7) of \cite{guslienko08}.  The approximate
expression for $\omega_G(0)$ obtained in the limit of small aspect
ratios $\beta\ll1$ has the form $\omega_G(0)=5/(9\pi) 4 \pi \gamma M_s
\beta$ \cite{guslienko02}. The value of the saturation field $H_s$ can
also be determined as the crossing point of the dispersion curves
(\ref{uniform}) and (\ref{slavin}). It occurs at $H_s \simeq H_K + 2
\omega_G(0)/ \gamma$ found to be $\approx 6.1$~kOe for the smaller
disk, a result that is close to the experimentally measured value
shown in Fig.\ref{vortex}a.

\begin{figure}
\includegraphics[width=\columnwidth]{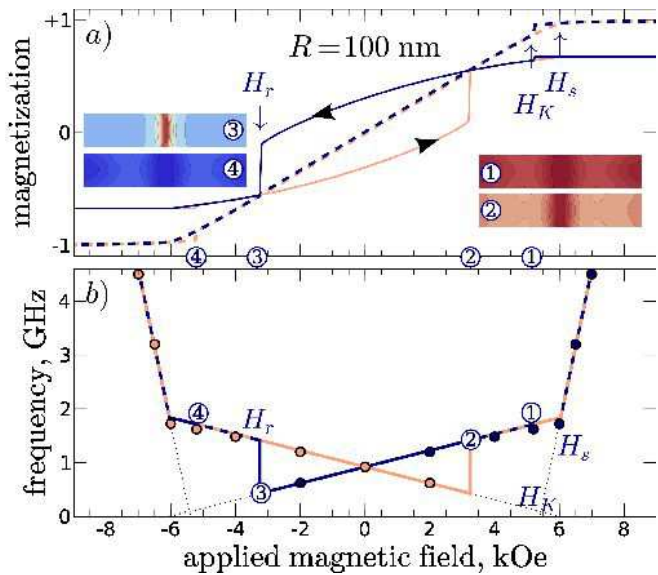}
\caption{(Color online)(a) Numerically calculated field dependence of
  the static magnetization of the NiMnSb disk (thickness $t=43.75$~nm,
  radius $R=100$~nm): the hysteretic behavior of the static
  magnetization, $M_z$, averaged over the disk volume (dashed loop)
  and over the vortex core (solid loop). The inserts show the spatial
  distribution of $M_z$ at four progressively decreasing values of
  $H$. (b) Resonance locus calculated by the 3D micromagnetic code
  (dots) along with the analytical model (solid lines).}
  \label{simu}
\end{figure}

The theoretical curves calculated using Eq.~(\ref{slavin}) are shown
by solid lines in Fig.\ref{vortex}, and it is clear from this figure
that Eq.~(\ref{slavin}) gives a good quantitative description of the
experimental data for two different values of the disk aspect
ratio. This simple analytic theory does not account for the reversal
of the vortex core polarity observed experimentally at $H=H_r$ (see
Fig.(\ref{vortex})). Therefore, we performed micromagnetic simulations
to reproduce this experimentally observed effect. The results of these
simulations are presented in Fig.\ref{simu}.

Our micromagnetic numerical code \cite{boust04} calculates the stable
configuration of the magnetization vector in the nano-disk by solving
the Landau-Lifshitz (LL) equation in the time domain. A 3D mesh with a
cubic cell of size 3.125~nm gives a discrete representation of the
nano-disk magnetization.  The nano-disk dimensions were chosen as
$R=100$~nm, $t=43.75$~nm, while the values of the magnetic parameters
were taken from table \ref{ref}. The numerically calculated static
hysteresis loop of $M_z\left(H\right)$ (where the vertical component
of the static magnetization $M_z$ was averaged over the disk volume)
is shown in Fig.\ref{simu}a by a dashed line. Color images of $M_z$
distribution across the radial section of the disk taken at four
points corresponding to progressively decreasing values of the
perpendicular bias magnetic field $H$ are shown as inserts in
Fig.\ref{simu}a. Although the vortex core switching is not seen in
this \emph{averaged} static hysteresis loop, it is clearly revealed
when the averaging of the static $M_z$ component is done over the
spatial region occupied by the vortex core ($\simeq 15$~nm), as
demonstrated by the solid lines in Fig.\ref{simu}a, where the core
polarity switching appears as a jump at $|H_{r}|= 3.25$~kOe.

It should be noted, that the vortex core reversal in the negative
perpendicular bias field was studied analytically and numerically in
\cite{thiaville03} and was attributed to the formation of a Bloch
point. Experimentally this effect was studied on an array of magnetic
dots in \cite{okuno02}. The magnitudes of the core reversal fields
obtained in \cite{thiaville03,okuno02} are consistent with
corresponding magnitudes obtained in our experiment Fig.\ref{vortex}
and numerical modeling Fig.\ref{simu}a.

The \emph{dynamic} hysteresis loop (i.e. the dependence of the
gyrotropic mode frequency on the perpendicular bias magnetic field
$H$) shown in Fig.\ref{simu}b was numerically calculated using a
second code developed by the authors, which computes the full dynamic
susceptibility tensor $\chi''$ from linearization of the LL equation
around the equilibrium configuration \cite{boust04}. It clearly
demonstrates the effects of the bias-field-induced dynamic bistability
\cite{giesen05} in the field interval $|H|< H_r$. Additional numerical
calculations are in progress to study the details of the gyrotropic
frequency behavior in the vicinity of the points $H=\pm H_s$ , where
the vortex is formed, and $H=\pm H_r$, where the reversal of the
vortex core polarity takes place.

This research was partially supported by the ANR PNANO06-0235 from
France, by the European Grants DynaMax (FP6-IST 033749) and Master
(NMP-FP7 212257), by the MURI Grant No. W911NF-04-1-0247 from the
U.S. Army Research Office, by the Contract W56HZV-08-P-L605 from the
U.S. Army TARDEC, RDECOM, and by the Grant No. ECCS-0653901 from the
U.S. National Science Foundation. K.G. acknowledges support by the
Ikerbasque Science Foundation.

\end{document}